\documentclass[aps,prl,preprint]{revtex4-1}
\usepackage{epsfig}
\usepackage{bbold}
\usepackage{amsmath}
\usepackage{mathrsfs}
\usepackage{makeidx}
\usepackage{enumerate}
\begin{document}
\author{R.J. Creswick}
\author{F.T. Avignone  III}
\affiliation{Department of Physics and Astronomy\\
University of South Carolina\\
Columbia, SC 29208}
\title{ALP Search Using Precessing Light in a Magnetized Fabry Perot Cavity }
\begin{abstract}
In this paper we outline an experiment to detect the conversion of photons to axion-like particles (ALPs) in a strong magnetic field. We show that by modulating the polarization of the light passing through a Fabry-Perot cavity so that it effectively precesses at the modulation frequency, a signal is  produced that is proportional to the square, as opposed to the fourth power, of the ALP-photon coupling constant. Assuming shot noise to be the dominant source of noise, we estimate that this approach is  sensitive to ALPs with masses less than $10^{-4} \text{eV}$ and couplings on the order of $g_{a\gamma}>1.6\times 10^{-11} \text{GeV}^{-1}$ with a 10m, 10 T magnet, and $g_{a\gamma}>1.6\times 10^{-12} \text{GeV}^{-1}$  with a 100 m magnet as envisaged by ALPs-IIc. ALPs with these properties have been invoked to explain the apparent transparency of the extragalactic background light (EBL) to ultra high-energy gamma rays emitted by BLAZARs.
\end{abstract}

\maketitle
\section*{Introduction}
A number of extensions of the Standard Model of Elementary Particle Physics predict the existence of a hidden sector of particles that interact with the visible sector extremely weakly. One such particle is the axion, a pseudo-scalar Goldstone boson arising from the spontaneous breaking of the global U(1) symmetry introduced into the strong interaction Lagrangian by Peccei and Quinn to solve the strong CP-problem [1-3]. The Peccei-Quinn (PQ) axion has a fixed relationship between its mass, $m_a$, in eV and the coupling to photons, $g_{a\gamma}$, in $\text{GeV}^{-1}$
\begin{equation}
g_{a\gamma}=10^{-9}\Bigl(0.203\frac{E}{N} -0.39\Bigr)m_a   \text{  GeV}^{-1}
\end{equation}
 The parameters $E$ and $N$ are the model-dependent electromagnetic and color anomalies. Characteristic values are $E/N=0$ in the KSVZ model \cite{K,SVZ} and $E/N=8/3$ in the DFSZ model \cite{DFS,Z}. Recent work \cite{light_axion,light_axion2} has extended this region to much lower axion masses. Other extensions of the Standard Model predict axion-like pseudoscalar bosons (ALPs) whose mass and interaction strength are independent. For a complete discussion of ALPs and their properties see the review by Jaeckel and Ringwald \cite{ALPS}.
 \par
ALPs can couple to the electromagnetic field, leptons and quarks. In this work we are concerned with the coupling of ALPs to the EM field, which is described by  a term in the Lagrangian of the form
\begin{equation}
\mathscr{L}_{a\gamma}
=g_{a\gamma}\boldsymbol{E} \cdot \boldsymbol{B}\phi
\label{eq:lint}
\end{equation}
where $\boldsymbol{E}$ and $\boldsymbol{B}$ are the electric and magnetic fields, and $\phi$ is the ALP field. 
This interaction gives rise to a process in which a photon can convert to an ALP in the presence of an external electric or magnetic field, analogous to the Primakoff effect in which a photon converts to a $\pi^0$ in the field of a nucleus, and the inverse Primakoff effect in which an ALP can convert to a photon in an external electromagnetic field. 
\begin{figure}[ht]
\includegraphics[width=90mm]{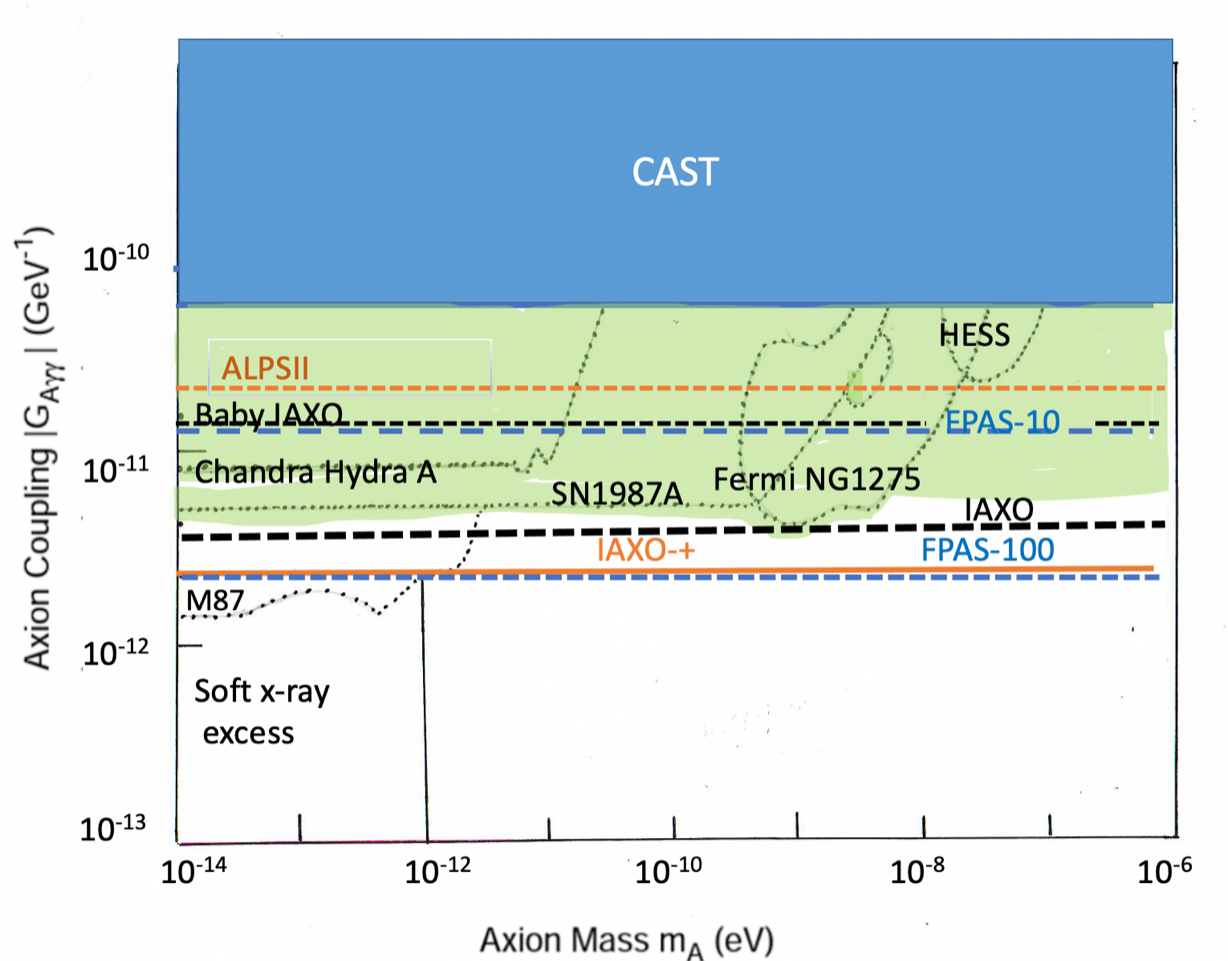}
\caption{ALP parameter space \cite{PDG} in the low-mass region relevant to extragalactic conversion of high-energy photons to ALPS.  The limits labeled FPAS-10 and FPAS-100 are projected sensitivities for the proposed experiment for L=10 and 100 m cavities and a 10 T magnet. Expected sensitivities for IAXO \cite{IAXO}, baby IAXO \cite{babyIAXO} and ALPs-II \cite{ALPS2}are also shown.}
\label{fig:exclusion}
\end{figure}
The hunt for ALPs takes place in the $m_a-g_{a\gamma}$ parameter space spanning many decades in both parameters, as can be seen in figure \ref{fig:exclusion}. 
Large parts of the ALP parameter space have been excluded by astrophysical arguments which are reviewed by Raffelt \cite{Raffelt} and Reynolds et. al. \cite{Reynolds}.  In addition to purely astrophysical arguments, experiments like ADMX \cite{ADMX} and CAST \cite{CAST} are designed to directly detect ALPs created in astrophysical  sources. 
\begin{figure}[ht]
\includegraphics[width=50mm]{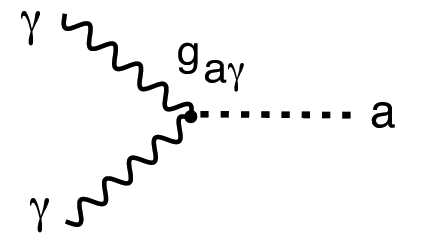}
\centering
\caption{Process of axion to photon decay looked for in haloscopes like ADMX.}
\label{fig:axion_to_photon}
\end{figure}
The Feynman diagram for the conversion of an ALP to a photon by the inverse Primakoff effect is shown in figure \ref{fig:axion_to_photon}. 
A review of experimental bounds is \cite{Experimental Search}
In addition to a solution of the strong CP problem, ALPs are a candidate for dark matter. ADMX \cite{ADMX} looks for photons created by the conversion of relic axions through the inverse Primakoff effect. Assuming PQ axions account for all the dark matter in the galaxy, their density is fixed by the axion mass. Axions in the galactic halo would have very little kinetic energy so the resonance of the ADMX cavity is tuned to a frequency equal to the axion mass. The rate of conversion is proportional to $g_{a\gamma}^2$.   
\par
CAST \cite{CAST} looks for ALPs created in the solar core and converted to photons in the magnetic field of the helioscope. Since the solar ALP flux is proportional to $g_{a\gamma}^2$ and the rate of reconversion to a photon is also proportional to $g_{a\gamma}^2$ the signal in the CAST helioscope is proportional to $g_{a\gamma}^4$. The CAST  experiment has excluded ALPs with coupling larger then $6.6\times 10^{-11}\text{GeV}^{-1}$ and masses less than about $10^{-2}$eV.   

 \par
Finally,  there are ALP searches that are completely laboratory-based. Among these are \lq Shining Light Through the Wall'  (SLW) experiments like ALPS-I \cite{ALPSI} in which a photon is first converted to an axion in a magnetized Fabry-Perot cavity. The axion then passes through an opaque barrier and then through a second magnetic field where it can  reconvert to a photon. The generation/regeneration process is inherently proportional to $g_{a\gamma}^4$. In ALPs-II \cite{ALPS2}, a second `regerneration' cavity is included afer the opaque barrier which greatly enhances the probability of reconversion to a photon. 
 Another type of laboratory-based experiment  that makes use of the polarization degrees of freedom of light and the form of the coupling to ALPs in \eqref{eq:lint} is exemplified by PVLAS \cite{PVLAS1, PVLAS2, PVLAS3}. The coupling between ALPs and photons produces both a birefringence and dichroism in a magnetized cavity which are proportional to $g_{a\gamma}^2$. 
\par
The birefringence is detected as an ellipticity of the light that is initially linearly polarized and dichroism as a rotation in the direction of the polarization. By using heterodyne detection techniques, a signal can be extracted at twice the modulation frequency. In order to eliminate a signal due to birefringence of the cavity even in the absence of a magnetic field the magnetic field is slowly rotated relative to the initial direction of polarization of the light. 
\section{\textbf{F}abry-\textbf{P}erot \textbf{A}LP \textbf{S}earch: FPAS}
In this paper we introduce a modification of the approach taken by PVLAS. Rather than rotating the magnetic field while keeping the polarization of the light fixed, we keep the direction of the magnetic field fixed and use an electro-optical modulator (EOM) to cause the polarization of the light in the cavity to precess with period $2\pi/\Omega$. The modulation frequency of the polarization, $\Omega$, is fixed by the free spectral range of the cavity. By matching the periodicity of the modulation with the free spectral range of the cavity we ensure that when the carrier frequency of the laser is critically coupled to the cavity,  the side-bands produced by modulation will also be on resonance. Depending on the length of the cavity this is in the range of 10-100 MHz as opposed to 10-20 Hz in PVLAS. 
\par
\begin{figure}[ht]
\includegraphics[width=90mm]{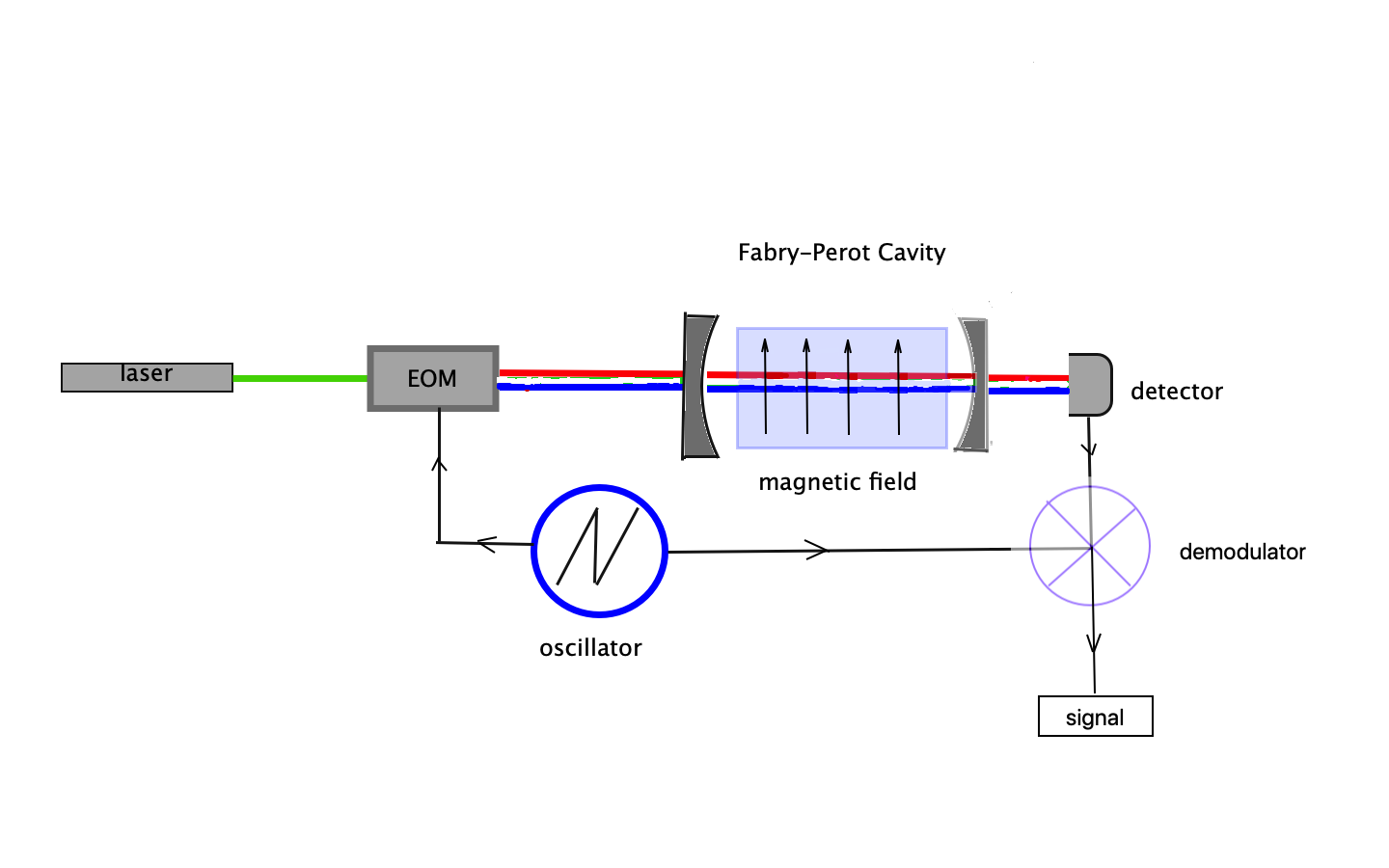}
\caption{Schematic diagram of the FPAS experiment.}
\label{fig:FPAS}
\end{figure}

Figure \ref{fig:FPAS} shows the basic elements of FPAS, many of which it shares with other laboratory-based experiments, especially PVLAS. An important distinction between FPAS and PVLAS is that in PVLAS the modulation frequency is comparable to the width of the resonance, so the first sidebands overlap significantly with the central resonance whereas in FPAS the sidebands overlap not with the central resonance of the cavity but with adjacent resonances.
\par
Given the coupling in \eqref{eq:lint}, light whose polarization is parallel to a magnetic field can convert to ALPs, but if the polarization is a right angles to the field the light will propagate as in free space.
In FPAS light initially polarized in the $\hat{\boldsymbol{x}}$-direction (parallel to the magnetic field) is made to precess at the frequency $\Omega$ by passing it through an electro-optical modulator (EOM) oriented so that the optical axes of the EOM crystal are
\begin{equation}
\begin{split} 
\hat{\boldsymbol{e}}_o&=(\hat{\boldsymbol{x}}+\hat{\boldsymbol{y}})/\sqrt{2}\\
\hat{\boldsymbol{e}}_e&=(\hat{\boldsymbol{x}}-\hat{\boldsymbol{y}})/\sqrt{2}
\end{split}
\end{equation}
The EOM is driven by an applied electric field at frequency $\Omega$ producing a phase difference $\Gamma(t)$ between the ordinary and extraordinary components of the light with period $T=2\pi/\Omega$. The electric field of the light that emerges from the EOM and enters the Fabry-Perot cavity is
\begin{equation}
\boldsymbol{E}_{in}(t)=E_0 e^{i(\alpha(t)-\omega t)}\bigl[\cos \Gamma(t )~\hat{\boldsymbol{x}}+i\sin\Gamma(t) ~\hat{\boldsymbol{y}} ]
\end{equation}
where $\omega$ is the frequency of the laser, $\alpha(t)$ is an over-all time dependent phase, and $\Gamma(t+T)=\Gamma(t)$ is periodic with period $T$. The detailed structure of the sidebands is determined by the waveform of the driving voltage, but in any case only involves frequencies $\omega_n=\omega+n\Omega$ for integer $n$. In the special case where $\Gamma(t)=\pi V_0/V_\pi \sin\Omega t$ \cite{sidebands}
\begin{equation}
\begin{split}
\boldsymbol{E}_{in}(t)&=E_0e^{i(\alpha(t)-\omega t)}\Bigl(\sum_{\text{n even}}J_n(\pi V_0/V_\pi)\cos n\Omega t~\hat{\boldsymbol{x}}\\
&+i\sum_{\text{n odd}}J_n(\pi V_0/V_\pi)\sin n\Omega t~\hat{\boldsymbol{y}}\Bigr)
\end{split}
\end{equation}
where $J_n(x)$ is the n-th Bessel function.
\par
The transmission amplitude for a Fabry-Perot (FP) cavity is 
\begin{equation}
\tau=\frac{t_1t_2e^{i\phi}}{1-r_1r_2e^{2i\phi}}
\end{equation}
where $r_1,t_1$ and $r_2, t_2$ are the reflection and transmission amplitudes for the two mirrors, and $\phi=\omega L/c$ is the phase change of the light in traversing the length of the cavity. 
\par
The resonances of the cavity are equally spaced in frequency at $\omega_n=\omega_0+n\Delta \omega$ where $\Delta \omega= \pi c/L$ is the free spectral range.  At resonance the transmission coefficient is very close to one (n even) or minus one (n odd) , so by matching the modulation frequency $\Omega$ to $\Delta \omega$, the even sidebands at $\omega\pm 2k\Omega$ are transmitted essentially unchanged and the odd sidebands at $\omega\pm (2k+1)\Omega$ change sign.  
\par
In the absence of a magnetic field, and ignoring scattering and absorption in the cavity, the polarization of the light that emerges from the cavity is in the state
\begin{equation}
\boldsymbol{E}_{out}(t)=
E_0e^{i[\alpha(t)-\omega(t-L/c)]}\Bigl[\cos \Gamma(t)\hat{\boldsymbol{x}} 
-i\sin\Gamma(t)\hat{\boldsymbol{y}}\Bigr]
\end{equation}
and the intensity of the light is exactly the same as the input intensity
\begin{equation}
|E|_{out}^2(t)=|E|_{in}^2(t)=E_0^2
\end{equation}
In the presence of a magnetic field $\boldsymbol{B}=B_0\hat{\boldsymbol{x}}$ the coupling to ALPs modifies the output field so that
\begin{equation}
\boldsymbol{E}_{out}(t)=
E_0e^{i[\alpha(t)-\omega( t-L/c)]}\Bigl[e^{i\theta}\sqrt{1-P_{\gamma\rightarrow a}}\cos \Gamma(t) \hat{\boldsymbol{x}}\\
-i\sin\Gamma(t) \hat{\boldsymbol{y}}\Bigr]
\end{equation}
where $P_{\gamma\rightarrow a}$ is the probability of conversion of a photon to an ALP in the cavity \cite{Sikivie} and is responsible for ALP-induced dichroism. 
\begin{equation}
\label{pconvert}
P_{\gamma\rightarrow a}=\frac{2\mathscr{F}}{\pi} \Bigl(\frac {g_{a\gamma} B_0 L}{2}\Bigr)^2\Bigr(\frac{2\sin qL/2}{qL}\Bigl)^2
\end{equation}
 where $\mathscr{F}$ is the finesse of the cavity and $q$ is the momentum difference between the photon and ALP. 
\par
The phase $\theta$ is of order $g_{a\gamma}^2$ and produces ALP-induced birefringence. 
The output of the FP cavity is  
 \begin{equation}
|E_{\text{out}}|^2=E_0^2\Bigl(1-\frac{P_{\gamma\rightarrow a}}{2}\Bigr)-\frac{E_0^2 P_{\gamma\rightarrow a}}{2}\cos2\Gamma( t)
\label{intensity}
\end{equation}
The intensity consists of a DC component and a time dependent component proportional to $P_{\gamma\rightarrow a}$, which in turn is proportional to $g_{a\gamma}^2$. This is the advantage of FPAS over ALP searches whose signals are proportional of $g_{a\gamma}^4$. 

\par
\section{Signal and  Sensitivity}
If we assume an ideal photodetector that produces a single electron for each photon, in the limit ${P_{\gamma\rightarrow a}\ll 1}$ the current out of the detector is 
\begin{equation}
I(t)=I_0(t)\Bigl(1-\frac{P_{\gamma\rightarrow a}}{2}\cos2\Gamma(t)\Bigr)
\end{equation}
where $I_0(t)$ the current in the absence of a magnetic field. We assume the shot noise in $I_0(t)$ is governed by a Poisson distribution with constant mean
\begin{equation}
\langle I_0(t)\rangle=I_0
\end{equation}
and is delta-correlated,
\begin{equation}
\langle \delta I_0(t_1) \delta I_0(t_1)\rangle=e I_0\delta(t_1-t_2)  
\end{equation}
where $e$ is the charge of the electron.
\par
The Fourier components of the signal  averaged over a time $T$ are, for $n$ even,
\begin{equation}
S_n=\frac{1}{T} \int_0^T dt \langle I(t) \rangle\cos n\Omega t=-\frac{1}{4}I_0 C_n P_{\gamma\rightarrow a}
\label{signal}
\end{equation}
and zero for $n$ odd. The coefficient $C_n$ depends on the details of the driving voltage of the EOM. The angular brackets represent a statistical average over the Poisson distribution. 
\par
We assume that the dominant source of noise is shot noise due to the random fluctuations in $I_0$. By increasing the power of the laser we can ensure that shot noise dominates other well-known sources of noise like Johnson noise and noise in the photodetector. The PVLAS group \cite{PVLAS3, Milotti} has identified `birefringence noise' associated with thermal fluctuations in the mirrors of the FP cavity.  This source of noise decreases with increasing modulation frequency and is currently almost two orders of magnitude greater than the shot-noise. The modulation frequency envisioned in FPAS is six orders of magnitude greater than that of PVLAS, and if the scaling behavior observed by PVLAS holds, our assumption that shot noise is dominant in FPAS should be valid.

\par
The RMS fluctuation in the signal is 
\begin{equation} 
\Delta S_{shot}=\sqrt{\frac{e I_0 }{2T}}
\end{equation}
and the signal-to-noise ratio is
\begin{equation}
SNR=\sqrt{\frac{I_0 T}{8e}}P_{\gamma\rightarrow a}
\end{equation}
Note that $I_0T/e$ is essentially the total number of photons that pass through the cavity during the time $T$. Assuming one electron per photon 
$$\frac{I_0}{e}=\frac{P_{laser}}{\hbar \omega}$$
where $P_{laser}$ is the power input into the cavity by the laser and $\omega$ is the frequency of the laser.
\par
The power in the cavity is related to the input power of the laser and the finesse by 
\begin{equation}
P_{\text{cav}}=\frac{\mathscr{F}}{\pi} P_{laser}
\end{equation}
and is limited by the laser induced damage threshold (LIDT) of the mirrors, which we take to be 750 kW. For a finesse of $5\times10^5$ this limits the power of the laser to 4.7W.
The finesse of the cavity also enters into the probability of photon-to-ALP conversion,
\begin{equation}
P_{\gamma\rightarrow a}=\frac{2\mathscr{F}}{\pi} P^{(1)}_{\gamma\rightarrow a}
\end{equation}
where $P^{(1)}_{\gamma\rightarrow a}$ is the probability of conversion without the FP cavity. 
\par
The SNR is maximized by increasing the finesse subject to constraint imposed by the LIDT. Taking this into account the SNR is 
\begin{equation}
SNR=\sqrt{\frac{P_{LIDT} \mathscr{F} T}{2\pi \hbar\omega}} P^{(1)}_{\gamma\rightarrow a}
\end{equation}
With light with wavelength 1064 nm, a 10 m cavity with finesse of $5 \times10^5$, and  one year of exposure, we find
$$SNR=3.18\times 10^{18}P^{(1)}_{\gamma\rightarrow a}$$
The probability to convert a photon to an ALP in a single pass assuming  $B=10$T, and $L=10$m in the low mass limit $m_ac^2/\hbar \omega\ll1$ is, with $g_{a\gamma}$ in $\text{GeV}^{-1}$,  
\begin{equation}
P_{\gamma\rightarrow a}^{(1)}=2.43\times 10^{3} g_{a\gamma}^2
\end{equation}
giving the signal to noise ratio
\begin{equation}
SNR=7.73\times10^{21} g_{a\gamma}^2
\end{equation}
Setting this equal to one leads to an estimate of the two-sigma sensitivity 
\begin{equation}
g_{a\gamma}>1.6\times10^{-11}\text{GeV}^{-1}
\end{equation}
and would cover the model space for ALPs of masses less than about $10^{-4}$ eV. 
This bound is about 4 times better than the current bound of $6.6\times 10^{-11}\text {GeV}^{-1}$ set by CAST  and more than three orders of magnitude better than ALPS-I. 
\par
Next-generation versions of ALPS (ALPS-II) and CAST (IAXO) are in the planning stages. By making use of the existing infrastructure of the HERA accelerator, ALPS-II \cite{ALPS2} is estimated to reach a sensitivity on the order a $10^{-11} \text{GeV}^{-1}$ with a 100m long magnet. IAXO \cite{IAXO} is designed to reach a sensitivity on the order of $10^{-12} \text{GeV}^{-1}$. With a 100m magnet as envisioned for ALPS-IIb FPAS could reach a sensitivity of $1.6\times10^{-12} \text{GeV}^{-1}$, which is very competitive with other next-generation ALP searches.

\section{ALPs and the Opacity the Extragalactic Background Light to Ultra-High Energy Photons}
Ultra-High Energy (UHE) gamma rays produced in distant Active Galactic Nuclei (AGNs) with energies on the order of a TeV  should be absorbed by pair creation with extragalactic background photons (EBL) \cite{EBL}. Nevertheless gamma rays in the TeV range associated with AGNs have been detected by Imaging Atmospheric Cherenkov Telescopes (IACTs), for example MAGIC \cite{MAGIC} and VERITAS \cite{VERITAS}.
\par
A possible explanation for these unexpected UHE photons  is described in recent papers by Galanti et al.\cite{BLAZAR, BLAZAR2}. 
Their proposed explanation is that UHE gamma rays produced in distant AGNs convert to ALPs in the strong magnetic field of the AGN. They then traverse the long distance to the Milky Way undisturbed until they enter the magnetic field of our galaxy where they reconvert into photons through the inverse process. Theoretical models of photon to ALP conversion in AGNs  and reconversion in the magnetic field of the galaxy favor the range of couplings  $10^{-12} \text{ GeV}^{-1}\le g_{a\gamma}\le 10^{-10} \text{ GeV}^{-1}$ and ALP masses on the order of $10^{-9} \text{eV}<m_a<10^{-7} \text{eV}$. This region is indicated in grey in figure \ref{fig:exclusion}, and provides an added motivation for next generation experiments like IAXO \cite{IAXO}, ALPS II\cite{ALPSII} and FPAS as shown in figure \ref{fig:exclusion}. Recent analysis of the interaction of ALPs with matter in the solar core \cite{solar core supression} suggests that the flux of solar ALPs may be reduced, relaxing the CAST bound of $g_{a\gamma}<0.66\times 10^{-10}\text{GeV}^{-1}$. 
\par
The direct observation of ALPs with these properties would be conclusive evidence for the proposed ALP-gamma oscillation mechanism that allows UHE gamma rays to be observed by IACTs. As shown by the curve labeled FPAS-100 in figure \ref{fig:exclusion}, FPAS will be able to probe this region of the ALP parameter space.
\section{Conclusions}
We have shown that introducing light whose polarization \lq precesses' relative to the direction of the fixed magnetic field in a high-finesse Fabry Perot cavity causes dichroism modulated at twice the frequency of the precession and proportional to $g_{a\gamma}^2$. Assuming shot noise the dominant source of noise, we estimate that a 10m, 10T cavity could set a bound of $g_{a\gamma}<1.6\times10^{-11} \text{GeV}^{-1}$, and with a 100 m 10 T cavity a bound an order of magnitude lower. This purely laboratory experiment will explore a significant part of the ALP parameter space, and in particular the region preferred by explanations of the pair production puzzle in which very high energy photons are able to traverse intergalactic distances without absorption by pair production with the extragalactic background light.
\end{document}